\documentclass[letterpaper, 10 pt, conference]{ieeeconf}
\IEEEoverridecommandlockouts    
\usepackage{multirow}
\usepackage{lipsum}
\usepackage{amsmath}
\usepackage{amssymb}
\usepackage[ruled,vlined]{algorithm2e}
\usepackage{graphicx}
\graphicspath{{./figures/}}

\usepackage{booktabs}
\usepackage{siunitx}
\usepackage{tikz}
\usepackage{xcolor} 

\makeatletter
\let\NAT@parse\undefined
\makeatother

\usepackage{xurl} 
\usepackage[hidelinks]{hyperref}

\newcommand\copyrighttext{%
    \footnotesize \copyright{ }2026 IEEE. Personal use of this material is permitted. Permission from IEEE must be obtained for all other uses, in any current or future media, including reprinting/republishing this material for advertising or promotional purposes, creating new collective works, for resale or redistribution to servers or lists, or reuse of any copyrighted component of this work in other works.}
\newcommand\copyrightnotice{%
    \begin{tikzpicture}[remember picture,overlay]
    \node[anchor=south,yshift=15pt,xshift=0pt] at (current page.south) {\parbox{\dimexpr\textwidth-\fboxsep-\fboxrule\relax}{\copyrighttext}};
    \end{tikzpicture}%
}

\title{\LARGE \bf
Cloud-Native Operation of Roadside Infrastructure \\Enabling Demand-Driven Collective Perception via V2X
}

\author{
	\parbox{\textwidth}{%
		\centering
        Lukas Zanger, Fabian Thomsen, Guido Linden, Jean-Pierre Busch, Lennart Reiher, and Lutz Eckstein
	}%
  \thanks{\newline All authors are with the Institute for Automotive Engineering (ika), RWTH Aachen University, Germany. {\tt\footnotesize \href{mailto:lukas.zanger@ika.rwth-aachen.de}{\{first.last\}@ika.rwth-aachen.de}}}%
}

\begin{document}
	
\maketitle
\bstctlcite{IEEEexample:BSTcontrol}
\thispagestyle{empty}
\pagestyle{empty}
\copyrightnotice
\vspace{-0.8\baselineskip}

\begin{abstract}
Intelligent roadside infrastructure is a key enabler for cooperative intelligent transport systems (C-ITS), supporting vehicles equipped with automated driving systems (ADS), e.g., through enhanced environment perception. With a growing number and an expanding functional scope of roadside units, scalable and efficient operation becomes a challenge. This paper presents a cloud-native architecture for the operation of distributed roadside infrastructure based on a Kubernetes cluster spanning roadside units and a cloud server.
Building on this architecture, a demand-driven orchestration approach is implemented to dynamically deploy resource-intensive services only when required. As a representative use case, a V2X-based collective perception application is deployed on-demand when a connected vehicle is nearby.
The approach is validated in a real-world experiment in our test field in Aachen, demonstrating that the collective perception application starts in time for the vehicle to benefit from it.
Without any demand, the application remains inactive, reducing energy consumption, channel congestion, and hardware wear.
Beyond the primary evaluation, V2X recordings from the test field are analyzed to estimate the energy-saving potential of demand-driven operation. 
In summary, the results demonstrate the practical feasibility of cloud-native, demand-driven operation of roadside infrastructure and indicate its potential to improve scalability and (energy) efficiency in future C-ITS deployments.

\end{abstract}

\section{Introduction}
\label{sec:introduction}

The advanced development of V2X communication enables cooperative intelligent transport systems~(C-ITS), in which vehicles and roadside infrastructure exchange information, leading to improved traffic safety and efficiency.
Intelligent roadside units equipped with advanced sensors are a key enabler for vehicle-infrastructure cooperation, complementing onboard perception by extending the situational awareness beyond line-of-sight and sensor limitations.

Advanced applications for C-ITS, such as infrastructure-assisted perception relying on information shared among multiple units, necessitate coordination and data exchange with each other and with edge or cloud servers.
Hence, the growing deployment and expanding functional scope of intelligent roadside infrastructure units~\cite{roadside_infrastructure_survey} require coordinated rather than isolated operation.
Software needs to be updated over-the-air, centrally monitored, and consistently rolled out or reconfigured across distributed roadside units in a coordinated manner.
In the context of software operation, paradigms like containerization, microservice architectures, and container orchestration enable automated deployment, configuration, and lifecycle management of software components.
Said paradigms are defined by the Cloud Native Computing Foundation (CNCF)~\cite{cloud_native_computing_foundation} as \textit{cloud-native techniques}. 
They are mature and widely adopted in conventional cloud environments, offering potential for scalable deployment of large numbers of roadside units at municipal scale.
Cloud-native techniques provide the foundation for automated cluster-wide updates, observability, and scalable software management for roadside infrastructure.

\begin{figure}[!t]
    \centering
    \includegraphics[width=0.8\linewidth]{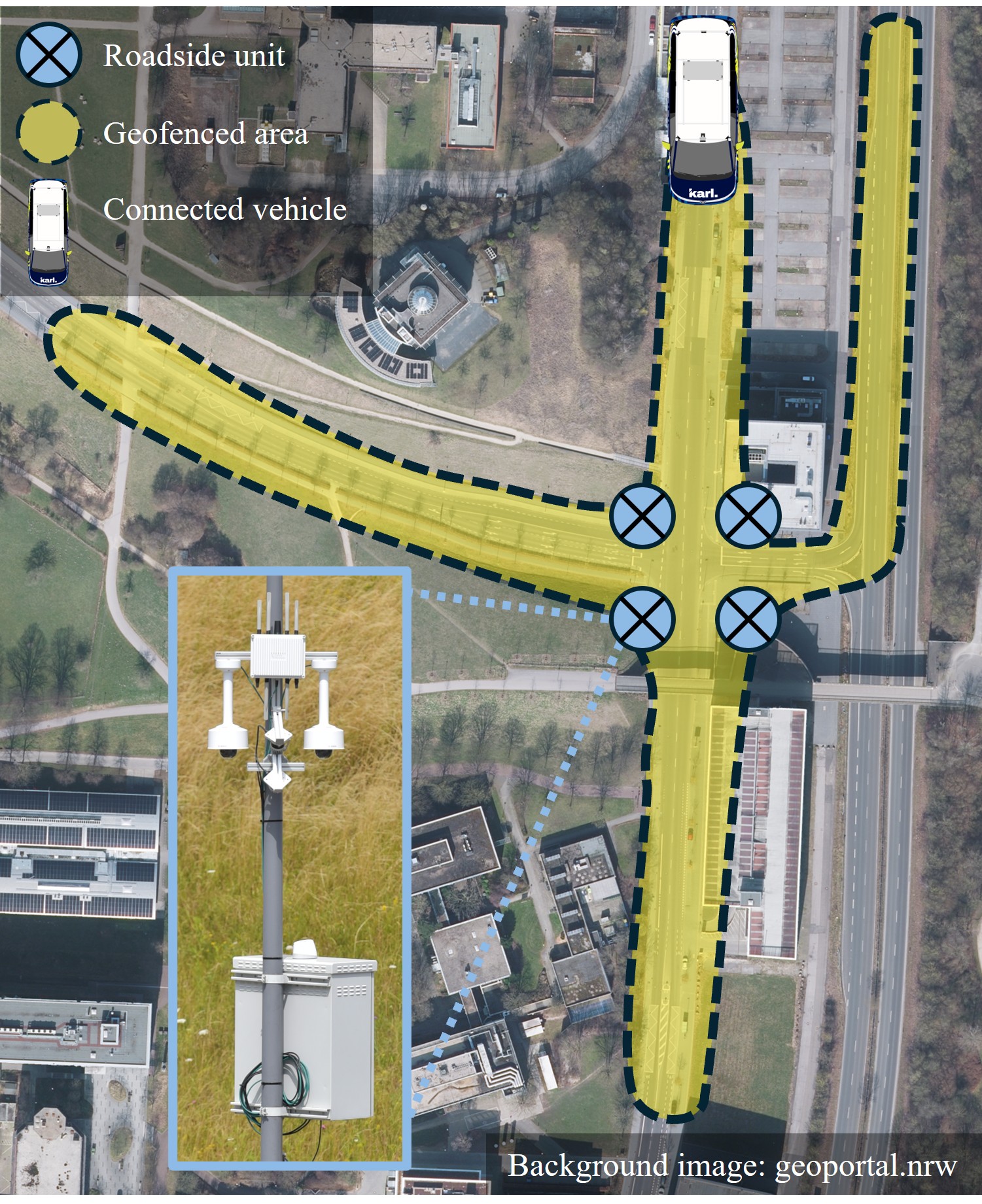}
    \vspace{-3mm}
    \caption{When a connected vehicle enters a geofenced area, services enabling collective perception are deployed to the roadside infrastructure units. As a result, ETSI CPMs are generated and broadcast to the vehicle via ITS-G5.}
    \vspace{-5mm}
    \label{fig:geo_fence_rita}
\end{figure}
C-ITS applications that involve, e.g., machine learning-based perception or planning algorithms may be computationally demanding. Consequently, they may only be deployed if a recipient is present, that is, when an actual demand exists. 
Such demand may involve requesting an external supporting service or triggering data recording for later use.
Without demand, continuous deployment would unnecessarily occupy computational resources and increase energy consumption. A demand-driven strategy reduces hardware utilization and operational costs over the lifecycle of the infrastructure units.

In this paper, we present a cloud-native architecture for the operation of intelligent roadside units in the \textit{Roadside Infrastructure Testfield Aachen} (\textit{RITA}) located at a road intersection (see Fig.~\ref{fig:geo_fence_rita}). 
We use the open-source platform Kubernetes (K8s), the de-facto standard for deploying scalable distributed systems~\cite{kubernetes_up_and_running}, as a container orchestrator.
The cloud-native operation of our test field (e.g., the usage of containerized microservices, centralized monitoring, and dynamic cluster-wide deployment of microservices) allows us to investigate the feasibility of a demand-driven orchestration strategy in a real-world setting. 
We examine a V2X-based collective perception application as a representative use case, in which software components are deployed based on demands resulting from the presence of a connected vehicle transmitting its position information.
We present a real-world experimental validation of the proposed approach in our test field,  providing a detailed latency analysis. 

In summary, we make the following main contributions:
\begin{itemize}
  \item Presentation of a cloud-native architecture for the operation of roadside infrastructure, implementing a distributed K8s cluster across multiple roadside units and a central server to enable scalable deployment, lifecycle management, and monitoring of C-ITS applications.
  \item Application of a demand-driven orchestration approach for automatically deploying resource-intensive services enabling a V2X-based collective perception application.
  \item Real-world experimental validation in our test field, demonstrating on-demand deployment of collective perception and providing a detailed latency analysis.
  \item Quantitative estimation of energy-saving potential of demand-driven operation, derived from V2X recordings.
\end{itemize}

\section{Fundamentals and Related Work}
\label{sec:related_work}

The approach presented in this work builds upon cloud-native techniques that involve paradigms such as containerization, microservice architectures, and container orchestration.
In \textit{microservices} architectures, applications are structured as a collection of individual services that can be developed, deployed, and scaled independently~\cite{cncf_microservices_architecture}.
Through \textit{containerization}, application code including its dependencies is packaged into lightweight executable units~\cite{cncf_containerization}.
\textit{Container orchestration} refers to the lifecycle management of containerized applications in distributed computing environments~\cite{cncf_container_orchestration}.
The open-source platform Kubernetes, which has become the de-facto standard for container orchestration~\cite{kubernetes_up_and_running}, enables a wide range of operational tasks, including application deployment, reconfiguration, scaling, self-healing mechanisms, and monitoring.
Containerized software is scheduled through Kubernetes across multiple nodes within a cluster, where a node represents a physical or virtual machine.
A cluster denotes a collection of computers collaborating to achieve a common goal~\cite{cncf_cluster}.
Kubernetes is made available through so-called distributions. A distribution, e.g., \textit{K3s}~\cite{k3s}, is a packaged and customized version of Kubernetes that streamlines the installation, configuration, and operation in specific environments.

The demand-driven vehicle-infrastructure cooperation investigated in this work relies on V2X communication.
In the context of C-ITS, evolving standards for direct V2X communication, such as ITS-G5~\cite{ETSI_EN_303_797_V2_1_1} based on IEEE 802.11p, provide low-latency and reliable data exchange between vehicles and infrastructure.
The European Telecommunications Standards Institute (ETSI) has standardized a set of V2X message formats to create a common understanding between C-ITS entities.
The Cooperative Awareness Message (CAM)~\cite{ETSI_EN_302_637_2_V1_4_1} and the Collective Perception Message (CPM)~\cite{ETSI_TS_103_324_V2_1_1} are used in this work.
CAMs are periodically broadcast by C-ITS entities to share state information, such as their current position. CPMs are used to share information about perceived objects in the environment, enabling collective perception among, e.g., vehicles and roadside infrastructure.
The software package stack developed by the authors of~\cite{kueppersV2AIXMultiModalRealWorld2024} is used in this work to facilitate the usage of ETSI ITS messages within the Robot Operating System (ROS\,2)~\cite{ros2}.

The authors of~\cite{roadside_infrastructure_survey} highlight the increasing development stages and the diversity of sensors of intelligent roadside infrastructure in C-ITS over the past years.
They emphasize the potential of capturing traffic data and the real-time distribution of digital twins.
In the context of roadside infrastructure for C-ITS, our work contributes to the research on the operation of intelligent roadside infrastructure units by investigating the application of cloud-native techniques, specifically Kubernetes-based orchestration.
Recent literature, e.g.,~\cite{anujharishkumarchaudhariCloudNativeInfrastructurePowering2025} and \cite{dengLeveragingPublicCloud2025}, highlights cloud-native technology as a key enabler for autonomous and connected mobility systems, emphasizing containerization, Kubernetes-based orchestration, and edge cloud distribution to achieve scalable and flexible processing of vehicle and traffic data and to support the requirements of real-time connected-vehicle applications.
In particular, the authors of~\cite{anujharishkumarchaudhariCloudNativeInfrastructurePowering2025} identify lightweight Kubernetes distributions, such as K3s, as suitable for resource-constrained edge environments, enabling real-time processing capabilities for autonomous vehicles.
Furthermore, the authors of~\cite{duDependabilityAwareCoordinationInfrastructureSide2025} highlight Kubernetes as a valid foundation for operating roadside infrastructure. They propose a dependability-aware coordination layer on top of Kubernetes to detect faults and maintain service continuity.

Our work contributes to the practical applicability of cloud-native techniques in the context of C-ITS. We investigate cloud-native operation of roadside infrastructure, employing a demand-driven orchestration approach to selectively deploy resource-intensive perception services in response to V2X-triggered events. The demand-driven orchestration approach is built on an application management framework presented in one of our previous works~\cite{application_management_cits}.

\section{Cloud Native Operation of Roadside Infrastructure}

The experimental setup employed in this work comprises four sensor-equipped \textit{stationary roadside ITS station units} (\textit{sRISUs}) that are part of our \textit{Roadside Infrastructure Testfield Aachen} (\textit{RITA}).
The notion of an \textit{ITS station unit} follows the definition provided by ETSI, which distinguishes, among other entities, between vehicle ITS stations and roadside ITS stations. Those are integral components of ETSI's intelligent transport systems (ITS) reference architecture~\cite{ETSI_EN_302_665_2010, ETSI_EN_302_636_3_V1_2_1}.

\begin{figure*}[htbp]
  \centering
  \includegraphics[width=0.9\linewidth]{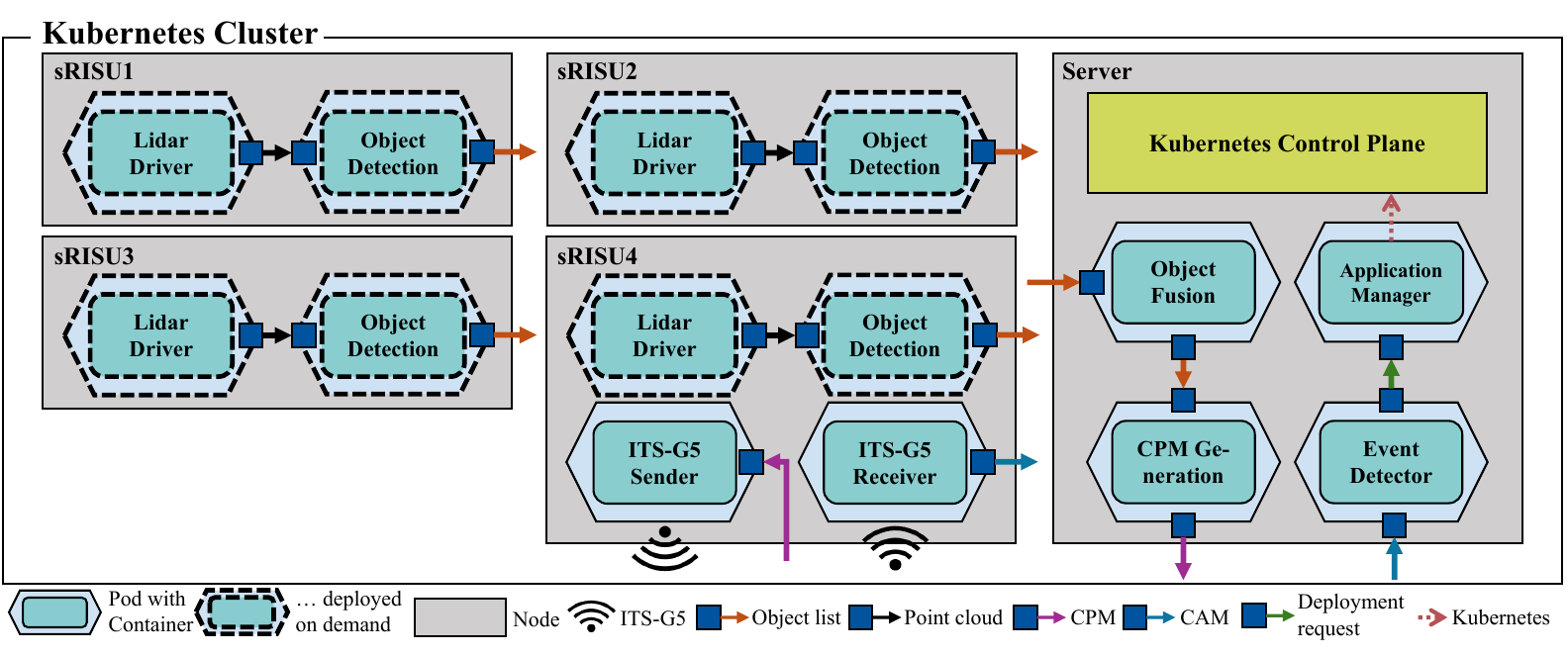}
  \vspace{-3mm}
  \caption{Kubernetes cluster comprising four sRISU nodes and one server node, collectively hosting the collective perception application. On each sRISU, lidar drivers and object detection services run in dedicated Kubernetes pods and are deployed on demand via the application management framework. One sRISU additionally hosts the ITS-G5 sender and receiver for CAM reception and CPM transmission. The server node runs the Kubernetes control plane and the components of the application management framework as well as services for object fusion and CPM generation. For clarity, auxiliary services (e.g., point cloud fusion) are omitted in the figure. Data transmission between nodes is indicated by arrows.}
  \label{fig:kubernetes_cluster}
  \vspace{-5mm}
\end{figure*}
The four sRISUs are located at a road intersection (see Fig.~\ref{fig:geo_fence_rita}) at Campus Melaten in Aachen, Germany. They are positioned with overlapping fields of view. Each sRISU is equipped with two lidar sensors, two PTRZ cameras, a V2X communication unit, and a computing unit featuring a GPU.
Time synchronization across all sRISUs is achieved using a GNSS receiver and the Precision Time Protocol (PTP). The V2X communication unit supports direct short-range communication via ITS-G5 and cellular communication used for cloud connectivity.
The computing unit with its GPU is responsible for real-time processing of the sensor data and for executing machine learning-based perception algorithms.

\subsection{Architecture of Kubernetes Cluster}

The sRISUs individually equipped with dedicated computational hardware constitute a distributed cluster of compute units. The units are interconnected with each other and with a central server via a Virtual Private Network (VPN) established over 5G cellular communication links. To automate the deployment of software components across distributed compute units, a Kubernetes cluster is implemented as visualized in Fig.~\ref{fig:kubernetes_cluster}. By adopting Kubernetes as the orchestration platform, the architecture benefits from the mature ecosystem and its advantages (see Section~\ref{sec:related_work}).

K3s \cite{k3s} is selected as the Kubernetes distribution due to its lightweight, resource-efficient design and suitability for resource-constrained Internet of Things (IoT) environments. Its architecture distinguishes between server and agent nodes: Server nodes host the control plane and datastore, while agent nodes execute workloads orchestrated by the server.~\cite{k3s}

In the proposed setup, each sRISU operates as an agent node, whereas a physically separate server hosts the control plane. Workloads are centrally deployed and managed across all sRISUs from the server node. 
Kubernetes mechanisms for, e.g., configuration and storage management facilitate scalable integration of additional sRISUs.

Advanced applications spanning multiple sRISUs consist of containerized microservices distributed across different nodes of the cluster. Following Kubernetes’ declarative model, deployments are defined using YAML manifests stored in version-controlled Git repositories.
Helm~\cite{helm} is used to bundle and parameterize related manifests through templated charts and version-controlled values files. It enables flexible adaptation and configuration of Kubernetes manifests considering different sRISU profiles and application requirements.

Monitoring is realized using Grafana~\cite{grafana} and the Kubernetes Dashboard~\cite{kubernetes_dashboard}. The Kubernetes Dashboard enables basic management operations, including web-based terminal access to containers for debugging.
Grafana monitors node and application resource utilization and other system-level performance indicators.
Metrics and container logs are aggregated using Prometheus~\cite{prometheus} and Grafana Loki~\cite{loki}.
GPU telemetry from the sRISUs equipped with NVIDIA GPUs is collected via the NVIDIA DCGM Exporter~\cite{nvidia_dcgm}.

\subsection{Demand-Driven Orchestration Approach}
\label{sec:demand_driven_orchestration}

The cloud-native operation of our test field, which includes the usage of containerized microservices, centralized monitoring, and dynamic cluster-wide deployment of microservices through the container orchestrator Kubernetes, allows us to apply the demand-driven application management approach presented in one of our previous works~\cite{application_management_cits}.

A demand, as explained in~\cite{application_management_cits}, may involve a request for provisioning an external supporting application or a trigger for recording data for later use.
A supporting application may only be demanded at certain times or regions of interest.
In a C-ITS, a demand may arise, for example, when a connected vehicle approaches an intersection equipped with roadside infrastructure enabling collective perception.
The collective perception application is examined as a representative use case in this work and is described in more detail in Section~\ref{sec:collective_perception_application}.
The demand-driven application management approach according to~\cite{application_management_cits} employs an operator application comprising an \textit{event detector} and an \textit{application manager}.
The reference implementation of the application management framework published on GitHub~\cite{application_manager_github} is utilized and extended for the use case of this work. Both the application manager and the event detector are containerized and deployed on the server node of the Kubernetes cluster (see Fig.~\ref{fig:kubernetes_cluster}).

The event detector~\cite{reiherevent} continuously monitors incoming data streams, e.g., V2X messages, and applies analysis rules to detect specific events that indicate a demand for an application. If such an event is detected, the event detector sends a deployment request to the application manager. 
The application manager then interprets the detected demand and deploys the required software components by installing the corresponding Helm charts with the configured values files.

\subsection{Live Digital Twin Architecture}
\label{sec:live_twin_architecture}

The collective perception application as described in Section~\ref{sec:collective_perception_application} relies on what we refer to as a \textit{Live Digital Twin} architecture.
It involves a processing pipeline enabling the online fusion of object-level data perceived by all four sRISUs, providing a real-time digital representation of traffic.

In the architecture, all four sRISUs communicate with the server via ROS\,2 with Zenoh~\cite{zenoh} as the middleware implementation. The data flow is indicated in Fig.~\ref{fig:kubernetes_cluster}.
Raw data from two lidar sensors on each sRISU are fused into a single point cloud (one per sRISU). The object detection component on each sRISU detects objects in the fused point cloud and generates an object list. The object lists are sent from the sRISUs to the central server where they are fused using the methods described in~\cite{tracking2023}.
The fused object list is converted into a CPM and serialized into a UDP bitstring.
One of the sRISUs broadcasts the CPMs via ITS-G5.
Moreover, CAMs transmitted by V2X-capable traffic participants are received via ITS-G5 and provided as additional inputs to the object fusion.

\subsection{Collective Perception Application}
\label{sec:collective_perception_application}

In this work, we employ a V2X-based collective perception application as a representative use case to evaluate a demand-driven orchestration strategy.
In this application, when a CAM from a connected vehicle is received by the roadside infrastructure and the vehicle is detected to be approaching the infrastructure, the processing pipeline of the Live Digital Twin architecture is triggered, and CPMs are generated and broadcast.
The microservices of the application are illustrated in Fig.~\ref{fig:kubernetes_cluster}.
Some of the microservices of the Live Digital Twin architecture, in particular, the lidar sensor drivers and the object detection, are deployed to the sRISUs only when an actual demand for collective perception exists.
Such demand arises when a connected vehicle transmitting its position as a CAM via ITS-G5 approaches the intersection equipped with sRISUs.
The application management framework deploys the required software components to the sRISUs once a CAM from the connected vehicle arrives.

Two extensions within the application management framework (Section~\ref{sec:demand_driven_orchestration}) are required to realize the demand-driven collective perception application.
First, an analysis rule is implemented for the event detector. It continuously monitors incoming CAMs from connected vehicles and detects whether a vehicle is located in the vicinity of the intersection, e.g., within a geofenced area as illustrated in Fig.~\ref{fig:geo_fence_rita}. If so, a deployment request is sent to the application manager.
In the experiment described in Section~\ref{sec:experiment_setup}, we do not apply a geofence but simply wait for the first CAM received from the vehicle. Based on our experimental results, an optimized geofence could be determined for a later production deployment.

Second, a new application class is implemented within the application manager, containing logic to configure and deploy the sensor driver and object detection services on demand.

\section{Experimental Setup for Collective Perception}
\label{sec:experiment_setup}

The feasibility of the demand-driven orchestration approach for the collective perception application is evaluated through a real-world experiment.
It involves the RITA test field and our research vehicle \textit{karl.}\,\cite{busch2026karl} which is equipped with an \mbox{ITS-G5} onboard unit.
The CAMs sent by \textit{karl.}\,contain, besides the current position, a station ID that is pseudonymously assigned to the sender.
The V2X hardware of the sRISUs is able to receive these CAMs as soon as the distance between \textit{karl.}\ and the sRISUs is within the communication range of ITS-G5 which can be up to 800 meters~\cite{maglogiannisExperimentalV2XEvaluation2022},\,\cite{C2CCC2021CoChannel}.

The experimental use case involves the following steps:
\begin{itemize}
    \item \textit{karl.}\ approaches the RITA test field and periodically broadcasts CAMs containing a specific station ID.
    \item \textit{sRISU4} receives incoming CAMs. The stream of CAMs is subscribed by the event detector on the server. The event detector analyzes whether there are CAMs indicating the presence of a participant for collective perception.
    \item As soon as \textit{sRISU4} receives CAMs from \textit{karl.}\,, the event detector notifies the application manager to deploy the corresponding microservices on the sRISUs.
    \item After the startup of all services is accomplished, the data flow (see Fig.~\ref{fig:kubernetes_cluster}) is established. The object fusion receives input object lists from all sRISUs and fuses them. A CPM is sent back to \textit{sRISU4}, where it is broadcast via ITS-G5.
    \item Finally, the CPM is received by \textit{karl.}\,.
\end{itemize}

Three runs of the experiment are conducted. The intersection (see Fig.~\ref{fig:geo_fence_rita}) is approached by \textit{karl.}\ in one run from the north, in one from the west, and in one from the south.
We record both container logs and ROS 2 bags on \textit{karl.}\,, the sRISUs, and on the server. The recorded ROS 2 topics (e.g., CAMs, CPMs, object lists) and the container logs contain timestamps of the corresponding messages and events, which are used for the latency analysis (Section~\ref{sec:latency_analysis}).
Moreover, GPU power consumption of the sRISUs is recorded.

\section{Evaluation}
\label{sec:evaluation}

The data recorded during the real-world experiment allows us to evaluate the feasibility of the demand-driven orchestration approach by means of a latency analysis. 
Moreover, V2X recordings from the test field are analyzed to estimate the energy-saving potential of demand-driven operation.

\begin{figure}[htbp]
  \centering
  \includegraphics[width=1\linewidth]{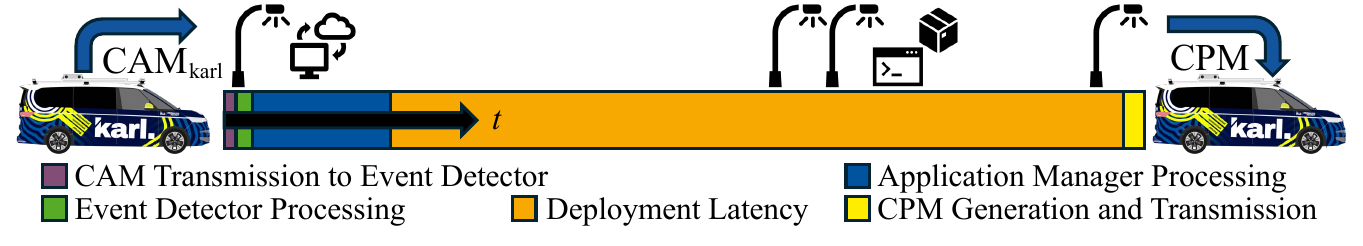}
  \vspace{-7mm}
  \caption{End-to-end latency is defined as the time between \textit{karl.}\ sending $\text{CAM}_\text{karl}$ and receiving the first CPM. It is composed of intermediate latency components as detailed in the subsection below.}
  \vspace{-6mm}
  \label{fig:latency_sketch}
\end{figure}

\subsection{Latency Analysis}
\label{sec:latency_analysis}
We examine the end-to-end latency of the collective perception application, which is composed of, e.g., communication latencies and latencies introduced by the demand-driven orchestration strategy. The end-to-end latency is the time elapsed between \textit{karl.}\ sending the message $\text{CAM}_\text{karl}$ and receiving the first CPM generated and broadcast by the sRISUs in response to $\text{CAM}_\text{karl}$.
Fig.~\ref{fig:latency_sketch} illustrates the latency components.
From all CAMs received by \textit{sRISU4}, we identify $\text{CAM}_\text{karl}$ by \textit{karl.}\,'s station ID. It is the first CAM received from \textit{karl.}\,. Upon arrival of $\text{CAM}_\text{karl}$, the orchestration process (see Sections~\ref{sec:demand_driven_orchestration} and~\ref{sec:collective_perception_application}) is initiated.

\begin{table}[b]
\vspace{-4mm}
\caption{Measured latencies per experiment run}
\vspace{-2mm}
\centering
\begin{tabular}{@{}p{0.4cm} p{1.6cm} p{2.4cm} p{1.6cm} p{0.8cm}@{}}
\toprule
Exp. No. & End-to-End Latency [s] & Application Manager Processing [s] & Deployment Latency [s] & Other [s] \\
\midrule
1 & \textbf{12.346} & 1.854 &  9.843\footnotemark[1] & 0.650\footnotemark[1] \\
2 & \textbf{12.457} & 1.805 &  9.593\footnotemark[1] & 1.059\footnotemark[1] \\
3 & \textbf{12.602} & 1.737 & 10.279\footnotemark[1] & 0.586\footnotemark[1] \\
\bottomrule
\end{tabular}
\label{tab:e2e_latency}
\end{table}

The measured latencies are summarized in Table~\ref{tab:e2e_latency}.
The end-to-end latency can be measured solely based on data recorded at \textit{karl.}\ and is therefore independent of time synchronization. In contrast, the evaluation of intermediate latencies relies on the analysis of data recorded across multiple entities (\textit{karl.}\,, sRISUs, and the central server) with potentially imperfect time synchronization.
Timestamps referring to certain events of the orchestration process are extracted from the recorded ROS~2 bags and the container logs and are used to calculate the latencies.
An excerpt from the intermediate latencies, which add up to the end-to-end latency, is provided in Table~\ref{tab:e2e_latency} for each experiment run. They consist of the following components:\textcolor{white}{\footnote[1]{Due to imperfect clock synchronization between the server and the sRISUs, the reported timestamps are subject to an uncertainty of \qty{\pm 20}{\milli\second}.}}
\begin{itemize}
  \item Deployment latency, for which a detailed breakdown is provided in Table~\ref{tab:deployment_latency} referring to the first experiment run, representing the majority of the the latency introduced by the demand-driven orchestration strategy;
  \item the application manager processing time, which is the duration between the application manager receiving the deployment request and the execution of the Helm release installation (including the download from a registry, processing and configuration of the Helm charts; excluding the actual startup of the K8s Pods);
  \item and other latencies, which include communication latencies (e.g., transmission of CAMs and CPMs via \mbox{ITS-G5}, RMW Zenoh communication between sRISUs and server) and processing latencies (e.g., event detection, object fusion, CPM and UDP bitstring generation).
\end{itemize}

\begin{table}[ht]
\caption{Breakdown of deployment latency for experiment no. 1 measured on \textit{sRISU1} (Note that the steps no. 4-5 and 6-7 run in parallel to steps no. 1-2)}
\vspace{-2mm}
\centering
\begin{tabular}{@{}p{0.1cm} p{6.3cm} p{1.4cm}@{}}
\toprule
& Deployment Latency Component & Duration [s] \\
\midrule
1 & Creation of K8s Pod with service \textit{Object Detection} & 6.099\footnotemark[1] \\
2 & First detection callback after startup of \textit{Object Detection} & 0.370 \\
3 & First object list generated after first detection callback & 3.374 \\
$\sum$ & Sum & \textbf{9.843}\footnotemark[1] \\
\midrule
4 & Creation of K8s Pod with service \textit{Lidar Driver 1} & 4.162\footnotemark[1] \\
5 & First packets received by \textit{Lidar Driver 1} after startup & 0.630 \\
\midrule
6 & Creation of K8s Pod with service \textit{Lidar Driver 2} & 4.151\footnotemark[1] \\
7 & First packets received by \textit{Lidar Driver 2} after startup & 0.655 \\
\bottomrule
\end{tabular}
\vspace{-6mm}
\label{tab:deployment_latency}
\end{table}

The end-to-end latencies exhibit consistency across the three experiment runs, indicating a stable system response under the test conditions.
A component-wise analysis (Table~\ref{tab:e2e_latency}) shows that the deployment latency dominates the overall delay ($\sim$\,80\%).
The breakdown of the deployment latency (Table~\ref{tab:deployment_latency}) reveals that the Kubernetes Pod creation dominates the deployment latency, indicating that cold-start effects at the container runtime level constitute the primary bottleneck.
The results further indicate that cold-start overhead is service-dependent, as the object detection service exhibits a higher startup time than the lidar driver services.
Mitigation strategies for cold-start overhead, such as optimizing container images or improving ROS 2 node lifecycle management, could reduce deployment latency.

Based on the measured end-to-end latency, we derive the distance within which a connected vehicle (CV), namely \textit{karl.}\ in our experiment, must be located for the deployment to start in time, so that the CPM is received by the CV while it still has not reached the intersection of interest yet. This enables the derivation of, e.g., the dimensions of a potential geofence.
Assuming the CV is moving with a speed of \qty{50}{km/h}, which is valid for an urban scenario, the CV covers a distance of approximately 180 meters during the end-to-end latency of around 13 seconds.
Adding a planning horizon of ten seconds, during which an ADS-equipped vehicle could plan its maneuvers based on the received CPM, the distance increases to approximately 300 meters.
Hence, the CV must be within a distance of approximately 300 meters from the intersection for the deployment to start in time.
A geofence, as exemplarily illustrated in Fig.~\ref{fig:geo_fence_rita}, could be defined accordingly. Such a geofence should not be modeled as a circular area, but should be adapted to the road layout and the relevant approaching lanes or road segments.
The calculated distance of 300 meters, taking into account an urban scenario, is within the communication range of ITS-G5 (up to 800 m \cite{maglogiannisExperimentalV2XEvaluation2022},\,\cite{C2CCC2021CoChannel}), even with a safety margin.
This emphasizes the practical feasibility of the demand-driven orchestration approach for the collective perception application triggered by V2X messages sent via ITS-G5.

In the scope of the latency analysis, we additionally examine the ITS-G5 communication latency more closely.
For one experiment run with \textit{karl.}\ approaching the intersection from the north, we first measure the latency of the CAM transmission from \textit{karl.}\ to \textit{sRISU4}.
Second, we measure the latency of the CPM transmission from \textit{sRISU4} to \textit{karl.}\,. The latencies are averaged over all recorded messages. The results are provided in Table~\ref{tab:its_g5_latency}.
The observed difference between CAM and CPM latencies may partially originate from residual clock synchronization inaccuracies between \textit{karl.}\ and \textit{sRISU4}, potentially on the order of 1--2\,milliseconds.

\begin{table}[t]
\caption{Averaged ITS-G5 communication latencies}
\vspace{-2mm}
\centering
\begin{tabular}{@{}lccccc@{}}
\toprule
& \multicolumn{5}{c}{Latency [ms]} \\
\cmidrule(l){2-6}
Transmission & Mean & Median & Std & Min & Max \\
\midrule
CAM (\textit{karl.} $\rightarrow$ \textit{sRISU4}) & 8.17 & 8.17  & 2.23 & 3.22 & 22.91 \\
CPM (\textit{sRISU4} $\rightarrow$ \textit{karl.}) & 5.25 & 5.15  & 2.29 & 0.19  & 11.79 \\
\bottomrule
\end{tabular}
\label{tab:its_g5_latency}
\vspace{-5mm}
\end{table}

\subsection{Analysis of V2X Recordings}

In the second part of the evaluation, we analyze V2X recordings collected in our test field to estimate the potential for power savings through demand-driven orchestration, under assumptions that are specific to our test field.
Through the V2X units of our test field, we record CAMs sent by \mbox{ITS-G5}-capable vehicles seven days in a row from Monday until Sunday in February 2026. The recording contains 69610 CAM messages sent by 714 unique station IDs (vehicles).

\begin{figure}[b]
  \vspace{-5mm}
  \centering
  \includegraphics[width=\linewidth]{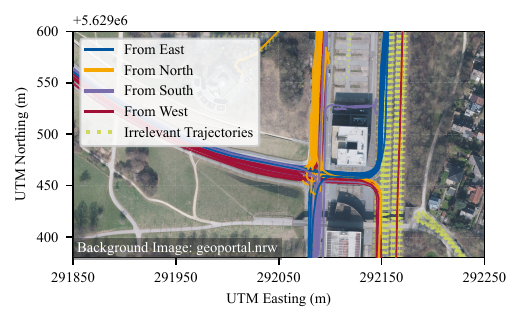}
  \vspace{-9mm}
  \caption{Trajectories are aggregated based on the recorded CAM messages. The colors represent different directions from which the vehicles have approached the intersection.}
  \label{fig:cam_trajectories_maneuver}
\end{figure}
First, we analyze the trajectories of the vehicles based on the position information and the station ID contained in the CAM messages. Vehicles (i.e., station IDs) that passed through the intersection are identified, as these would have triggered the demand-driven orchestration.
The resulting aggregated trajectories are shown in Fig.~\ref{fig:cam_trajectories_maneuver}. The trajectories are color-coded according to the respective route. The routes denote the directions from which the vehicles have approached the intersection.
The ``irrelevant'' trajectories (see Fig.~\ref{fig:cam_trajectories_maneuver}) are excluded from the subsequent analysis.

\begin{figure}[htbp]
  \centering
  \includegraphics[width=1\linewidth]{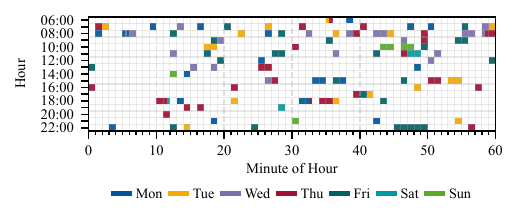}
  \vspace{-9mm}
  \caption{CAM occurrence matrix for the V2X recording: The blocks indicate the occurrence of CAM messages at minute-level resolution. Colors represent the individual days of the examined week. The time interval from 23:00 to 6:00 is omitted as no trajectory was detected as relevant during those hours.}
  \label{fig:time_matrix_day_singled}
  \vspace{-5mm}
\end{figure}
The CAMs associated with the remaining trajectories are analyzed to derive the temporal distribution of received CAM messages per day, resolved at minute-level granularity. 
Based on this distribution, we determine how frequently ITS-G5-capable vehicles would have triggered the demand-driven deployment and, consequently, benefitted from the collective perception application.
The temporal distribution is shown in Fig.~\ref{fig:time_matrix_day_singled}. The blocks in the matrix indicate the occurrence of CAM messages.
Periods without CAM occurrences indicate time intervals during which the collective perception application could have remained inactive to reduce power consumption. For these intervals, the potential power savings are quantified for the specific configuration of our test field in the following.
We only consider the power consumption of the GPUs of the sRISUs.
The on-demand object detection services run on the GPUs of the sRISUs in our setup.

We measure that the averaged additional GPU power consumption on one sRISU is approximately \qty{45}{\watt} when the collective perception application is running compared to when it is not running.
With four sRISUs in our test field, this results in an additional power consumption of \qty{180}{\watt}.
Per minute, this amounts to an additional energy consumption of \qty{3}{Wh}.
On average over the seven days, 18~minutes with CAM occurrence are counted per day.
Considering a buffer of one minute after the CAM occurrence, the collective perception application would have been active for 33 minutes per day on average ($\sim$\,2\% of the day).
This reflects the moderate traffic density at the intersection and may not be representative for heavily trafficked urban environments.
According to the 33~minutes, the application could have remained inactive for 1407 minutes per average day.
A continuous activation of the application would have resulted in an avoidable additional energy consumption of approximately \qty{4.22}{kWh} per day across the four sRISUs under these conditions.
Extrapolated over one year, this corresponds to a potential energy saving of approximately \qty{1500}{kWh} for the test field.
As a rough reference, this amount is on the order of the annual electricity consumption of a single-person household in Germany~\cite{destatis_stromverbrauch_haushalte}.
Since the additional consumption scales approximately linearly with the number of deployed units, larger municipal deployments would proportionally increase the energy demand.
For example, a deployment of 100 comparable sRISUs under similar traffic conditions would result in an additional annual energy demand of approximately \qty{37.5}{MWh}, highlighting the relevance of energy-aware activation strategies for large-scale roadside infrastructure.

It should be noted that these derivations are subject to assumptions specific to our test field. Those include the local traffic density, on which we are unable to provide quantitative information, and the use of power-demanding GPU hardware. 
Also, the current market penetration of the ITS-G5 standard in production vehicles influences the observed triggering frequency and must be considered when interpreting the results.

\section{Conclusion}

This work demonstrates that cloud-native techniques can be systematically transferred to the operation of intelligent roadside infrastructure and that they provide a viable foundation for scalable, energy-aware C-ITS deployments. By implementing a distributed Kubernetes cluster spanning multiple roadside infrastructure units and a central server, we establish an operational architecture that enables automated deployment, lifecycle management, monitoring, and coordinated execution of distributed services.

Building on this foundation, we apply a demand-driven orchestration approach in which resource-intensive software components are deployed upon detection of actual demand, triggered via V2X communication in our setup. 
The real-world evaluation in our test field with four roadside infrastructure units confirms the technical feasibility of the proposed approach, as demonstrated through a V2X-based collective perception application serving as a representative use case in this work.
We measure end-to-end deployment latencies in a real-world experiment, demonstrating that the derived geofence dimensions for an urban scenario are within the communication range of ITS-G5. 
Beyond functional feasibility, the analysis of week-long V2X recordings from our test field reveals an estimate for the energy-saving potential of the demand-driven approach for the collective perception application in our test field.
This indicates the relevance of energy-aware orchestration strategies for larger scale municipal deployments of intelligent roadside infrastructure.

In future work, we will extend the Kubernetes cluster by integrating additional roadside units at further locations. Moreover, the established architecture enables the investigation of additional on-demand V2X-based applications.

\section*{ACKNOWLEDGMENT}

This work was funded by the European Union under the Horizon Europe programme in the projects ``iEXODDUS'' (Grant Agreement No.~101146091) and ``SYNERGIES'' (Grant Agreement No.~101146542).

\bibliographystyle{IEEEtran}
\bibliography{root} 
	
\end{document}